\documentclass[]{relaxed_system_lab}

\usepackage[utf8]{inputenc}
\usepackage[T1]{fontenc}
\usepackage{geometry}
\usepackage{amsmath,amssymb}  %
\usepackage{charter}

\usepackage[toc,page,header]{appendix}

\usepackage{algorithm}
\usepackage[noend]{algorithmic}
\usepackage{array}
\usepackage{wrapfig}

\usepackage{amsmath}
\usepackage[utf8]{inputenc} 
\usepackage[T1]{fontenc}    
\usepackage{hyperref}       
\usepackage{url}            
\usepackage{booktabs}       
\usepackage{amsfonts}       
\usepackage[most]{tcolorbox}
\usepackage{nicefrac}       
\usepackage{microtype}      
\usepackage{xcolor} 
\usepackage{amssymb}
\usepackage{caption}
\usepackage{xspace}
\usepackage{multirow}
\usepackage{xcolor}
\usepackage{graphicx} 

\newcommand{\sys}{\textsc{Cascadia}\xspace}

\newcommand{\jyh}[1]{\textcolor{black}{#1}}
\ifdefined\final
\usepackage[disable]{todonotes}
\else
\usepackage[textsize=tiny]{todonotes}
\fi

\usepackage{natbib}
\usepackage{latexsym}

\usepackage{url}
\usepackage{amssymb}
\usepackage[utf8]{inputenc}
\usepackage{microtype}
\usepackage{booktabs}
\usepackage{pifont} 
\usepackage{multirow}
\usepackage{makecell}
\usepackage{paralist}
\usepackage{xspace}
\usepackage{color}
\usepackage{xcolor}
\usepackage{colortbl}
\usepackage{adjustbox}
\usepackage{hyperref} 
\usepackage[edges]{forest}
\usepackage{tikz} 
\usepackage{caption}
\usepackage{amsfonts}

\hypersetup{
    colorlinks,
    linkcolor={blue!80!black},
    citecolor={blue!80!black},
}
\tikzset{
    root/.style =             {align=center, text width=1cm, rounded corners=3pt, line width=0.3mm, fill=gray!10, draw=gray!80, font=\small},
    demographic/.style =         {align=center, text width=1.8cm, rounded corners=3pt, line width=0.3mm, fill=blue!10, draw=blue!80, font=\footnotesize},
    demographic_work/.style =    {align=center, text width=10cm, rounded corners=3pt, line width=0.3mm, fill=blue!10, draw=blue!0, font=\footnotesize},
    character/.style =         {align=center, text width=1.8cm, rounded corners=3pt, line width=0.3mm, fill=red!10, draw=red!80, font=\footnotesize},
    character_work/.style =    {align=center, text width=10cm, rounded corners=3pt, line width=0.3mm, fill=red!10, draw=red!0, font=\footnotesize},
    personalization/.style =           {align=center, text width=1.8cm, rounded corners=3pt, line width=0.3mm, fill=cyan!10, draw=cyan!80, font=\footnotesize},
    personalization_work/.style =      {align=center, text width=10cm, rounded corners=3pt, line width=0.3mm, fill=cyan!10, draw=cyan!0, font=\footnotesize},
    risk/.style =         {align=center, text width=1.8cm, rounded corners=3pt, line width=0.3mm, fill=orange!10, draw=orange!80, font=\footnotesize},
    risk_work/.style =    {align=center, text width=10cm, rounded corners=3pt, line width=0.3mm, fill=orange!10, draw=orange!0, font=\footnotesize},
}

\usepackage{CJK}

\newtcolorbox{promptbox}[1][]{
  enhanced,
  breakable,
  colback=promptboxlightgray,
  colframe=promptboxblue!30,
  arc=8pt,
  boxrule=0.5pt,
  left=12pt,
  right=12pt,
  top=8pt,
  bottom=8pt,
  fonttitle=\bfseries,
  fontupper=\linespread{1.2}\selectfont,
  title=#1
}

\title{\sys: An Efficient Cascade Serving System for Large Language Models}

\author{Youhe Jiang$^1$$^*$, Fangcheng Fu$^2$$^*$, Wanru Zhao$^3$$^*$, Stephan Rabanser$^4$, \\ Jintao Zhang$^5$, Nicholas D. Lane$^3$, Binhang Yuan$^1$$^\dagger$}

\affiliation{$^1$HKUST, $^2$Shanghai Jiaotong University, $^3$University of Cambridge, \\ $^4$Princeton University, $^5$Tsinghua University}

\contribution{$^*$Equal contribution, $^\dagger$Corresponding author}

\abstract{
Recent advances in large language models (LLMs) have intensified the need to deliver both rapid responses and high-quality outputs. More powerful models yield better results but incur higher inference latency, whereas smaller models are faster yet less capable. Recent work proposes balancing this latency–quality trade-off using model cascades, which route simpler queries to smaller models and more complex ones to larger models. However, enabling efficient cascade serving remains challenging. Current frameworks lack effective mechanisms for handling (i) the huge and varying resource demands of different LLMs, (ii) the inherent heterogeneity of LLM workloads, and (iii) the co-optimization of system deployment and routing strategy.
Motivated by these observations, we introduce \sys, a novel cascade serving framework designed explicitly to schedule request routing and deploy model cascades for fast, quality-preserving LLM serving. \sys employs a bi-level optimization method: at the deployment level, it uses a mixed-integer linear program to select resource allocations and parallelism strategies based on LLM information and workload characteristics; at the routing level, it applies a Chebyshev-guided method to iteratively co-optimize the routing strategy and the system deployment produced by the deployment level. Our extensive evaluation on diverse workload traces and different model cascades (DeepSeek and the Llama series) demonstrates that \sys significantly outperforms both single-model deployments and the state-of-the-art cascade serving baseline, achieving up to 4$\times$ (2.3$\times$ on average) tighter latency SLOs and up to 5$\times$ (2.4$\times$ on average) higher throughput while maintaining target answer quality.
}

\begin{document}

\maketitle

\section{Introduction}

\begin{wrapfigure}{r}{0.35\linewidth}
    \centering
    \vspace{-1em}
    \includegraphics[width=\linewidth]{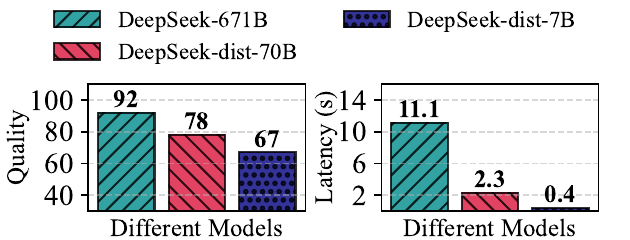}
    \vspace{-1.5em}
    \caption{\small{Average response quality and latencies of different DeepSeek models. Quality is judged by GPT-4o using the LLM-as-a-Judge framework~\citep{zheng2023judging}.}}
    \vspace{-1em}
    \label{fig:diffcascade}
\end{wrapfigure}

Large language models (LLMs) such as DeepSeek-R1~\citep{guo2025deepseek}, OpenAI o3~\citep{gpt4o}, Claude~\citep{claude3}, Gemini~\citep{reid2024gemini} and Llama-3~\citep{dubey2024llama} have demonstrated outstanding performance across a wide range of real-world applications (e.g., chatbots, healthcare and education)~\citep{jeon2023large,peng2023study,copilot}, largely influence human lives. However, serving LLMs can be costly~\citep{jiang2024hexgen,jiang2025hexgen,miao2024spotserve}, since significant computational resources (e.g., GPUs) are required to meet certain service demands, such as meeting certain latency deadlines (i.e., SLO attainment—the proportion of requests served within a specified response-time target) and generation throughput. In this paper, we explore an alternative solution that strategically utilizes model cascades to better balance the response latency and quality trade-offs inherent in LLM serving.

Cascade model serving refers to a serving architecture where multiple models of varying sizes and capabilities are arranged in a sequential pipeline, creating a hierarchy of models that process requests with increasing levels of sophistication~\citep{aggarwal2024automix,chenfrugalgpt,kossmann2024cascadeserve,kolawolerevisiting,lebovitz2023efficient,streeter2018approximation}. As shown in~\autoref{fig:diffcascade}, larger models typically provide higher response quality but also incur greater latency, which in turn leads to increased energy consumption and compute usage~\citep{samsi2023words}.
In this approach, incoming requests are initially handled by smaller, computationally efficient models that can rapidly process simpler requests. Only when these lightweight models determine that a request exceeds their capabilities or requires higher-quality responses does the system escalate the request to larger, more powerful models in the cascade. This progressive delegation mechanism enables service providers to optimize system performance by matching request complexity with appropriate model capacity, thereby significantly reducing computational costs while maintaining high-quality responses for complex request. Several recent studies have focused on optimizing LLM serving using model cascades~\citep{chenfrugalgpt,aggarwal2024automix,kossmann2024cascadeserve,guptalanguage,narasimhan2024faster}.

The cascade model serving architecture, which adaptively routes simpler and more complex requests to smaller and larger models, respectively, presents significant opportunities for optimizing the cost-efficiency of LLM serving. In this work, we focus specifically on the setting where service providers host and manage every model in the cascade themselves. However, effectively adapting this paradigm to LLM scenarios is much harder to implement than to propose, as we enumerate below:
\begin{itemize} %
\item \textbf{Model heterogeneity.} LLMs require large amounts of compute and memory, and different models have varying resource demands for efficient serving~\citep{duan2024muxserve}. With a fixed resource pool, suboptimal allocation across models in the cascade can degrade overall serving efficiency.


\item \textbf{Workload heterogeneity.} LLM workloads exhibit considerable heterogeneity~\citep{sun2024llumnix,zhenglmsys,zhaowildchat}. Models within the cascade often face incoming requests with varying characteristics (e.g., input/output lengths, arrival rates) and favor different deployment strategies (e.g., replication, parallel configuration), further adding complexity to optimal system deployment.

\item \textbf{Cascade-aware load balancing.} The request routing strategy directly impacts the system load of each model in the cascade. For instance, if more requests are routed to a particular model, its load increases; the resource allocation and deployment strategy for that model should then be adjusted to balance loads across all models. Consequently, the deployment of multiple models must be co‑optimized with the routing strategy to manage load across the cascade.
\end{itemize}
In order to overcome these challenges, we propose \sys, a novel cascade serving system that is optimized for LLM characteristics and that co‐optimizes the deployment of multiple models in the cascade together with the request routing strategy.
Our contributions are as follows:

\begin{itemize} 
\item \textbf{\underline{Contribution 1.}} We formulate cascade serving—covering system deployment and request routing—as a constrained optimization problem. To solve it, we propose a bi-level approach that jointly optimizes deployment and routing. The \emph{deployment} level uses mixed-integer linear programming (MILP) to determine the optimal deployment plan given a routing strategy, while the \emph{routing} level applies a \jyh{Chebyshev-guided} method to optimize routing, balancing latency and quality.

\item \textbf{\underline{Contribution 2.}} We implement \sys, an efficient cascade serving system tailored to LLMs. \sys enables an adaptive model cascade paradigm that allocates resources and routes requests across a hierarchy of model sizes (e.g., small, medium, and large), thereby balancing response latency and output quality. Within each cascade stage, \sys supports various parallelism strategies (e.g., tensor and pipeline parallelism), which allows it to automatically select the optimal strategy based on model size, incoming workload, and routing decisions.

\item \textbf{\underline{Contribution 3.}} We empirically evaluate \sys by comparing it to both single-model and existing cascade serving systems across a variety of scenarios, including diverse workload traces (e.g., coding and mathematics), different model cascades (DeepSeek and the Llama series), and multiple evaluation metrics (SLO attainment and throughput). The results show that, compared with state-of-the-art non-cascade and cascade solutions, \sys achieves up to 4$\times$ lower latency deadlines (2.3$\times$ on average) and boosts system throughput by up to 5$\times$ (2.4$\times$ on average).
\end{itemize}

\section{Preliminary and Related Work}
\textbf{LLM inference phases and workload heterogeneity.} There are two phases within LLM inference: \textit{prefill} and \textit{decoding}. During the prefill phase, the model processes the input prompt to compute the key-value (KV) cache and generates the first token in a single step. In contrast, the decoding phase uses the last generated token and the KV cache as inputs to generate subsequent tokens in a token-by-token manner. Generally, the prefill phase is compute-bound, while the decoding phase is memory-bound~\citep{patel2024splitwise,zhong2024distserve,agrawal2024taming,jiang2025demystifying,zhang2025efficient}. 
LLM inference workloads exhibit heterogeneity in input, output token lengths and request arrival rate, which is called \textit{workload heterogeneity}. For instance, conversation workloads (short input and long output lengths) typically require more memory resources to handle the memory-bound decoding phase, while coding workloads (long input and short output lengths) demand more compute resources to manage the compute-bound prefill phase. Therefore, appropriately allocating resources based on workload demands is critical for optimal performance~\citep{zhao2024blendserve,jiang2025thunderserve}.

\textbf{Cascade model inference.} 
Current LLMs come in various sizes and configurations, offering a broad spectrum of choices. Effectively leveraging this diversity can balance trade-offs between response latency and quality during inference. 
Recent efforts propose cascade model inference to utilize models of differing complexities~\citep{dekoninckunified,narasimhanfaster}. In such architectures, an input prompt is processed through increasingly complex models, using threshold-based routing that stops computation once a cheaper model produces a confident enough answer. For instance, FrugalGPT~\citep{chenfrugalgpt} employs a dynamic LLM cascade strategy that routes queries through progressively stronger models (e.g., GPT-3.5 → GPT-4) based on real-time difficulty estimation, optimizing cost-efficiency without sacrificing accuracy. Similarly, AutoMix~\citep{aggarwal2024automix} uses intelligent layer-wise token routing to dynamically allocate computation based on input difficulty. CascadeServe~\citep{kossmann2024cascadeserve} automates and optimizes end-to-end inference with cascades, adjusting model deployment and request routing based on real-time system loads. However, existing systems overlook key LLM-specific workload characteristics and neglect the importance of co-optimizing system deployment with request routing (i.e., system-algorithm co-design).

\begin{figure}[h] 
    \centering
    \includegraphics[width=\linewidth]{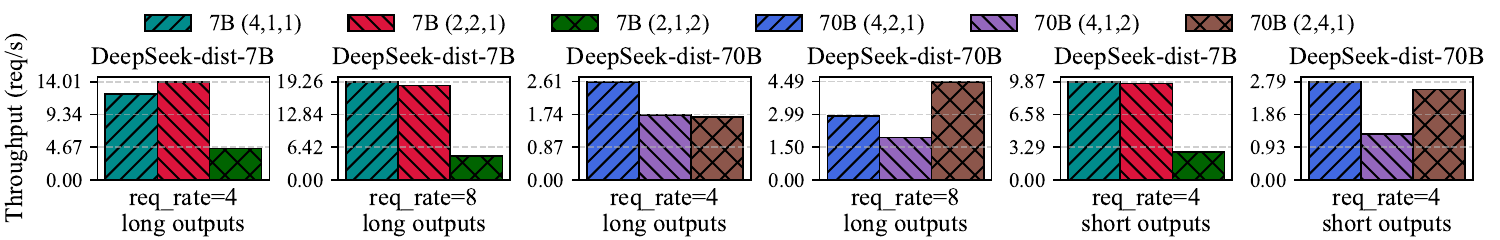}
    \caption{\small{Benchmarked performance of different parallelism strategies across different workloads and model sizes. Long and short outputs represent two different workloads with average output sequence length to be 512 and 1024; the three-element array represents the DP, TP, and PP degrees.}}
    \label{fig:parallel_impact}
\end{figure}

\textbf{Limitations of existing cascade serving systems.} We summarize the limitations of existing cascade serving systems: (\underline{\textbf{i}}) Ineffective resource allocation for different model types within a cascade. Different model types have distinct memory and computation resource needs. For example, DeepSeek-671B typically requires more allocated resources than DeepSeek-dist-70B due to its larger memory and computational demands. Current systems ignore the importance of adjusting resource allocation according to the needs of different model types,
leading to unbalanced system loads.
(\underline{\textbf{ii}}) Inadequate adaptation of parallelism strategies to varying workloads and model sizes. The optimal parallelism strategies vary across different workloads (e.g., different input and output request sequence lengths and request arrival rates) and model sizes. As shown in~\autoref{fig:parallel_impact}, choosing the optimal parallelism strategy can achieve up to 3$\times$ higher system throughput. Current systems do not optimize parallelism strategies according to specific workload and model size, resulting in degraded overall system performance. 
(\underline{\textbf{iii}}) Insufficient co-optimization between system deployment and routing strategy. The routing strategy decides the request portion processed by each model type within a cascade, which in turn determines the system loads for different model types. Existing systems neglect to adapt system deployment configurations based on routing outcomes, resulting in suboptimal resource usage. To address these challenges, a cascade serving system tailored for LLMs is necessary. Such a system must optimize end-to-end performance and ensure stringent SLO adherence.

\section{Scheduling Algorithm in \sys}

\subsection{Problem Formulation}
\label{subsec:problem formulation}
To optimize the cascade serving system under different LLM workloads \jyh{and user-specific requirements (e.g., system response quality requirements)}, the scheduling algorithm should determine two essential components: (\textbf{\underline{i}}) \textit{The model deployment plan}, which specifies the resource allocations and parallelism strategies for multiple model types (e.g., small, medium, large) within the cascade to minimize the system response latency (e.g., p95 latency—the response time threshold below which 95\% of all requests complete); and (\textbf{\underline{ii}}) \textit{the routing strategy}, which balances the trade-off between system response latency and quality to decide the appropriate model path for each incoming request. We term a solution addressing these two components as a \textit{cascade plan}. 


\begin{wrapfigure}{r}{0.48\textwidth}
{\small
\vspace{-10pt}
\captionsetup{type=algorithm}
\noindent\rule{\linewidth}{0.8pt}
\vspace{-19pt}
{\captionsetup{type=algorithm,
               font=footnotesize, labelfont=footnotesize, textfont=bf,
               labelfont=bf,
               justification=raggedright, singlelinecheck=false}%
 \captionof{algorithm}{Bi-level Scheduling Workflow}}
\hrule
\begin{algorithmic}[1]
\REQUIRE  
  $\theta_0$: initial routing strategy; \
  $\theta$: routing strategy; \
  $q_{\min}$: quality requirement; \ 
  $\tilde{\mathcal{I}}$: subsampled input workload; \ 
  $\mathcal{W}$: workload distribution; \ 
  $Q$: system response quality; \ 
  $N$: resource limit; \ 
  $\mathcal{D}$: deployment plan; \ 
  $L$: system response latency; \
  $J$: latency-quality score; \ 
  $K$: consecutive stable iterations to break
\ENSURE final routing strategy $\theta$ and deployment $\mathcal{D}$
\STATE $\theta \gets \theta_0$ \textcolor{green}{/* $\theta_0$ detailed in \S\ref{subsec:wt} */}
\WHILE{true}
  \STATE $(\mathcal{W},\, Q) \gets \text{derived~\footnotemark[1] from } (\theta,\, \tilde{\mathcal{I}})$ 
  \STATE \textcolor{green}{/* Optimize deployment (\S\ref{subsec:ilp}) */}
  \STATE $(\mathcal{D},\, L) \gets \text{DeploymentSolver}(\mathcal{W},\, N)$
  \STATE \textcolor{green}{/* Optimize routing strategy (\S\ref{subsec:wt}) */}
  \STATE $(\theta,\, J) \gets \text{RoutingSolver}(L,\, Q,\, q_{\min})$
  \STATE \textcolor{green}{/* Terminate upon convergence */}
  \IF{$J$ is stable for $K$ iters}
    \STATE \textbf{break}
  \ENDIF
\ENDWHILE
\RETURN $(\theta,\, \mathcal{D})$
\end{algorithmic}
\vspace{-8pt}\noindent\rule{\linewidth}{0.8pt}\vspace{-20pt}
}
\end{wrapfigure}

\footnotetext[1]{\jyh{Given $\theta$ and $\tilde{\mathcal{I}}$, $\mathcal{W}$ is derived by aggregating per-model routed requests (including arrival rates and sequence statistics), while $Q$ is derived by aggregating quality scores of accepted outputs across all models~\citep{chenfrugalgpt}.}
}

Note that the routing strategy determines the request distribution over different model types, which in turn dictates the optimal model deployment plan, while the model deployment plan defines the system response latency that feeds back into the routing decision. Given the interdependent and exponentially large search space, determining the optimal cascade plan is an NP-hard problem. To solve this problem, we adopt a bi-level optimization method that enables system–algorithm co-design, which is shown in Algorithm 1, and can be summarized as:

\begin{itemize}
    \item \textbf{MILP-based deployment solver}: Given the routing strategy, the deployment solver (\S\ref{subsec:ilp}) employs an mixed-integer linear programming (MILP) formulation to capture system resource constraints and compute the optimal deployment plan that minimizes system response latency.
    \item \jyh{\textbf{Chebyshev-guided routing solver}: Based on the system response latency generated from the deployment solver and the user-specific quality requirement, the routing solver (\S\ref{subsec:wt}) applies a Chebyshev-guided method to find the optimal routing strategy that optimizes system response latency with respect to the quality requirement.}
\end{itemize}


\subsection{MILP-Based Deployment Solver}
\label{subsec:ilp}

\begin{wrapfigure}{r}{0.5\linewidth}
    \centering
    \includegraphics[width=\linewidth]{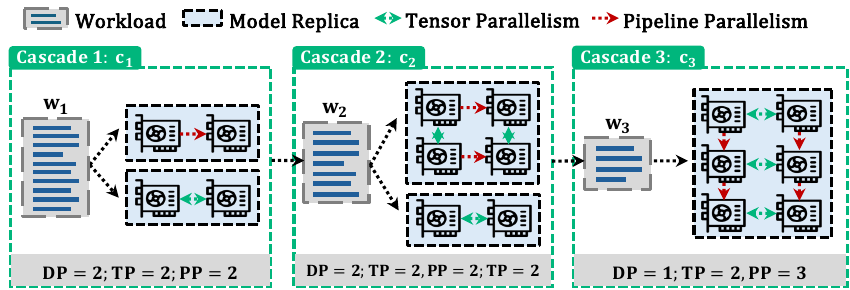}
    \caption{\small{Illustration of a model deployment plan.}}
    \label{fig:milp}
\end{wrapfigure}

\jyh{As shown in Algorithm 1, the routing strategy (obtained from routing solver) determines how many requests should be routed to each model in the cascade, thus determining the workload distribution among models. Given the \textbf{workload distribution} and \textbf{resource limit}, the deployment solver aims to determine the optimal \textbf{deployment plan}, which includes the resource allocation and parallelism strategies for models within cascades.} An example deployment plan is shown in~\autoref{fig:milp}. 

Assume a total of $N$ GPUs serve a model cascade with $C$ model types, $\{c_1, c_2, \dots, c_C\}$, where $c_i$ denotes the $i$-th model type. \jyh{The incoming workload information is denoted as $\mathcal{W}=\{w_1,w_2,\dots,w_C\}$, where each $w_i$ includes the distributions of input/output sequence lengths and the request arrival rate for the $i$-th model type.} We use $\mathcal{F}=\{f_1,f_2,\dots,f_C\}$ to denote the number of GPUs allocated per model, \jyh{the total allocation must not exceed the resource limit, i.e., $\sum_{i=1}^{C}f_i\leq N$. Given this setup, our deployment solver (\underline{\textbf{i}}) determines the parallelism strategy for each specific resource allocation $f_i$, and (\underline{\textbf{ii}}) uses an MILP to optimize the overall resource allocation $\mathcal{F}$.}


\textbf{Parallelism strategy search.} Given the workload information $w_i$ and a specific resource allocation $f_i$, this optimization determines the optimal parallelism strategy and computes the corresponding system response latency $l_i$ for the model type $i$. \sys provides three forms of parallelism: data parallelism (i.e., model replication, DP)~\citep{li2023alpaserve}, tensor model parallelism (TP)~\citep{shoeybi2019megatron}, and pipeline parallelism (PP)~\citep{huang2019gpipe}. Denoting the degrees of data, tensor, and pipeline parallelism for the model type by $\mathrm{dp}$, $\mathrm{tp}$, and $\mathrm{pp}$, any feasible parallelism strategy must satisfy the following resource constraint: $(\sum_{j=1}^{\mathrm{dp_i}} \mathrm{tp_{i,j}} \times \mathrm{pp_{i,j}}) \leq f_i$, i.e., one model type can be replicate into multiple replicas, each replica can have varied tensor and pipeline parallelism degrees, as shown in~\autoref{fig:milp}, the summation of different parallelism degrees should be less or equal than the total number of GPUs assigned. Based on the workload information $w_i$ and the resource allocation $f_i$, we iterate over all feasible parallelism combinations to select the strategy that minimizes the response latency $l_i$ for the model type $i$. The latency $l_i$ is computed using the simulator $\mathbf{Sim}(\cdot)$ as $l_i=\mathbf{Sim}(w_i, f_i)$~\footnote{We use the ETH EASL Scratchpad simulator~\citep{yao2023deltazip} to estimate system p95 latency from workload and resource allocation. \jyh{We show detailed simulator design (e.g., simulator inputs, batching strategy, queuing mechanism, parallelism strategy modeling) and evaluation in~\autoref{appendix:simu}.}}. \jyh{Note that the parallelism strategy optimization can be precomputed for all possible resource allocations $f$ to provide latency lookup tables for the MILP formulation.}

\textbf{MILP formulation for resource allocation optimization.} 
Our MILP problem formulation aims to minimize the maximum system response latency among all model types in the cascade. Let $L$ denote the maximum latency across all model types. We discretize the GPU allocations into candidate values $f \in \{1,2,\dots,N\}$. \jyh{For each model type $i$ and candidate allocation $f$, we use the precomputed latency table from the parallelism strategy optimization to obtain $l_i(f)$.} We then introduce binary assignment variables $x_{i,f}$, where $x_{i,f}=1$ if model type $i$ is assigned $f$ GPUs and $x_{i,f}=0$ otherwise, for all $i \in \{1,\dots,C\}$ and feasible $f$. The constraints of our MILP include: 
(\textbf{\underline{i}})~For each model type $i$, exactly one GPU allocation $f$ must be selected, i.e., $\sum_{f=1}^{N} x_{i,f} = 1, \forall\,i=1,\dots,C$; 
(\textbf{\underline{ii}})~the total number of GPUs assigned across all model types should be equal to the available GPUs $N$, i.e., $\sum_{i=1}^{C}\sum_{f=1}^{N} f\,x_{i,f} = N$; 
and (\textbf{\underline{iii}})~the maximum latency $L$ must be at least as large as the latency $l_i(f)$ corresponding to each selected allocation, i.e., $L \geq \sum_{f=1}^{N} l_i(f)\,x_{i,f}, \forall\,i=1,\dots,C$. 
We explicitly enforce variable domains and integrality constraints as follows: $x_{i,f} \in \{0,1\}, \forall\,i,f$ and $L \geq 0$. If certain GPU allocations $f$ are infeasible for specific model types—such as when the total memory of the allocated $f$ GPUs is less than the minimum memory required by the model type—we explicitly set $x_{i,f}=0$ for these allocation pairs. \jyh{Our objective is to minimize the maximum system response latency $L$, which serves as the input for the routing layer optimization.} 

\subsection{Chebyshev-guided Routing Solver}
\label{subsec:wt}

\begin{wrapfigure}{r}{0.35\linewidth}
    \centering
    \includegraphics[width=\linewidth]{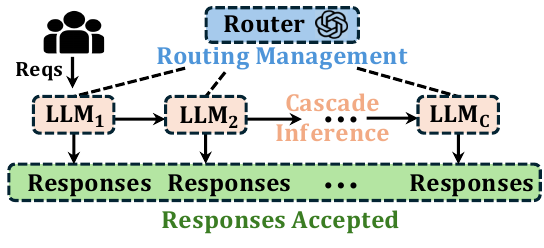}
    \caption{\small{Threshold-based cascade routing workflow. The router determines whether a request is accepted or forwarded to the next model type based on predefined thresholds.}}
    \label{fig:overflow}
\end{wrapfigure}

\jyh{As shown in Algorithm 1, the deployment plan (obtained from the deployment solver) determines the system response latency. Given the \textbf{system response latency} and \textbf{quality requirement}, the routing solver aims to optimize the \textbf{routing strategy} (i.e., co-optimize system latency and quality).}


\textbf{Thresholds tuning and request routing.} We adopt the threshold-based cascade routing workflow consistent with prior works~\citep{aggarwal2024automix,chenfrugalgpt} (\autoref{fig:overflow}).
Initially, every incoming request is sent to the first (smallest) model type $c_1$ in the cascade. A judger then evaluates the quality of the output responses from model types $c_1$ to $c_{C-1}$, and a set of thresholds $\mathcal{H}=\{h_1,h_2,\dots,h_{C-1}\}$ is defined to decide whether the requests at each model type should be accepted or forwarded to the next model type. In this framework, the routing strategy $\theta$ is directly determined by the thresholds $\mathcal{H}$, i.e., $\theta=\theta(\mathcal{H})$. Each routing strategy $\theta$ is associated with a system response latency $L(\theta)$ (determined by the deployment solver optimization) and quality $Q(\theta)$ (determined by the judger\jyh{~\footnote{\jyh{Analogous to~\citep{chenfrugalgpt}, we estimate $Q(\theta)$ by profiling a subsample of the input workload across all cascade models to obtain per-model quality score distributions. During scheduling, given any threshold vector $\mathcal{H}$ and the quality score distributions, we can determine which model's response would be accepted for each request under routing policy $\theta(\mathcal{H})$, then aggregate these final model scores to compute the overall system quality $Q(\theta)$.}}}). \jyh{Our routing solver uses a Chebyshev-guided method to optimize the routing strategy. We initialize the routing strategy $\theta_0$ as proportional routing, where the $i$-th model receives $1/i$ of requests.} 

\jyh{\textbf{Chebyshev-guided optimization for routing strategy.} Given the routing strategy $\theta$ 
and user-specified quality requirement $q_{\min}$, we employ the Chebyshev-guided method~\citep{steuer1983interactive} to minimize the system response latency $L(\theta)$ with respect to $q_{\min}$. 
First, we define a utopia point $z_1^*$ (all requests processed by the largest model $c_C$) and nadir point $z_2^*$ (all requests processed by the smallest model $c_1$) representing the best and worst achievable system response quality.
Then, for a given quality requirement $q_{\min}$, we minimize the system response latency subject to meeting the quality requirement by solving the single-objective penalty problem:
{\small
\[ \arg\min_{\theta} J(\theta) = \arg\min_{\theta} \left[ L(\theta) + \mu \max\{0,\; (q_{\min} - Q(\theta)) / (z_1^{*} - z_2^{*}) \} \right] \]
}%
where $J(\theta)$ represents the latency-quality score, $\mu>0$ is a penalty weight that enforces the quality constraint (for sufficiently large $\mu$, any minimizer of $J(\theta)$ satisfies $Q(\theta) \ge q_{\min}$), and $z_1^{*}$ and $z_2^{*}$ are used to normalize the quality shortfall so the penalty is dimensionless and well-conditioned across workloads. Note that our routing solver can also optimize system response quality under a user-specified latency requirement using a similar procedure, as detailed in~\autoref{appendix:routingsolver}.
}

\vspace{-2.0em}
\jyh{
\begin{tcolorbox}[colback=blue!5!white,colframe=blue!75!black]
\small
\textbf{Illustrative example for Chebyshev-guided optimization.} 
Assume the utopia and nadir points $z_1^*$ and $z_2^*$ equal $0.95$ and $0.75$. The user-specific quality requirement $q_{\min}$ is $0.90$ and the penalty weight $\mu$ is $100$. Consider a strategy $\theta_1$ with p95 latency $L(\theta_1) = 11.0$ s and overall quality $Q(\theta_1) = 0.88$. The normalized shortfall from the requirement is $(0.90 - 0.88)/(0.95 - 0.75) = 0.02/0.20 = 0.10$, yielding $J(\theta_1) = 11.0 + 100 \times 0.10 = 21.0$. 
Consider another strategy $\theta_2$ with latency $L(\theta_2) = 11.4$ s and quality $Q(\theta_2) = 0.91$, which results in $J(\theta_2) = 11.4$. Strategy $\theta_2$ is preferable under this setting due to its significantly lower objective value. 
Additionally, a higher-quality strategy $\theta_3$ with latency $L(\theta_3) = 12.2$ s and quality $Q(\theta_3) = 0.93$ yields $J(\theta_3) = 12.2$. Although both $\theta_2$ and $\theta_3$ satisfy the quality requirement $q_{\min}$, strategy $\theta_2$ is preferable since it achieves lower latency while meeting the constraint. This example demonstrates how the Chebyshev-guided method effectively penalizes infeasible solutions while optimizing system response latency.
\end{tcolorbox}
}

\jyh{\textbf{Putting them together.} In our bi-level optimization framework, the routing solver (i.e., Chebyshev-guided optimization) iteratively searches for the next $\theta$, invokes deployment solver (i.e., MILP optimization) to obtain the minimized system response latency $L(\theta)$, and then minimizes the objective function (i.e., $\arg\min_{\theta} J(\theta)$). Finally, an optimal routing strategy $\theta$ is selected that guarantees a minimal system response latency while fulfilling the quality requirement.}

\textbf{Impact of LLM workloads on optimal cascade plan selection.}
The characteristics of incoming LLM workloads strongly influence the selection of cascade plans. \jyh{This influence stems from two key factors}:
(\textbf{\underline{i}}) Request input/output length and arrival rate affect system response latency—longer sequences or higher loads increase compute demand, necessitating plan adjustments to balance latency and quality;
(\textbf{\underline{ii}}) Request complexity impacts system response quality—complex requests or difficult queries require larger models, necessitating plan adjustments to maintain quality while managing latency.
Therefore, our bi-level optimization framework considers both system performance (e.g., deployment solver) and algorithmic behavior (e.g., routing solver), enabling efficient, adaptive optimization across different incoming LLM workloads. \jyh{Additionally, our framework incorporates a re-scheduling mechanism to handle online fluctuating workloads, as detailed and tested in~\S\ref{sec:re-scheduling}.}

\jyh{The complete mathematical formulation for our bi-level optimization is provided in~\autoref{appendix:bilevel}.}

\section{Evaluation}





\subsection{Experimental Setup}
\label{sec:experimental_setup}

\textbf{Environments.} Our experiments are conducted on 4 GPU servers, where each server is equipped with 8 NVIDIA H100-80GB GPUs. Within each server, the GPUs are connected via NVLink with a bandwidth of 400GB/s, and the servers are connected via Inifiband with a bandwidth of 200GB/s.

\textbf{Model cascade construction.} We construct a model cascade using the DeepSeek series models for \sys, which are representative and popular open-source transformer models. Specifically, we use DeepSeek-dist-7B, DeepSeek-dist-70B (distilled version), and DeepSeek-671B AWQ with INT4 quantized weights~\citep{lin2024awq} as three model types within our system. We employ a GPT-4o (LLM-as-a-Judge)~\citep{zheng2023judging} as the judger mentioned in \S\ref{subsec:wt}, which assesses the output responses of each model type within the cascade and assigns scores between 0 and 100. \jyh{The judging overhead~\footnote{\jyh{The judger takes a Q\&A pair as input and outputs quality grades (1–2 tokens), resulting in significantly lower latency and cost than full request inference, typically requiring only 1–2s.}} is included in our experiments.}

\textbf{Baselines.} We compare \sys with two baselines: 
\begin{itemize}
    \item \textbf{Compare with stand-alone LLMs served by SGLang.} We compare \sys against stand-alone LLMs that are directly served on SGLang~\citep{zheng2024sglang} under various response quality constraints (e.g., 90, 85, 80, 70) to demonstrate the effectiveness of LLM serving with model cascades. For quality requirement of 90 and 85, we choose stand-alone DeepSeek-671B for comparison, and for quality reqirement of 80 and 70, we choose stand-alone DeepSeek-dist-70B for comparison. For fair comparison, we tune the parallelism strategy using our MILP algorithm mentioned in~\S\ref{subsec:ilp} for each of the stand-alone model and report the best values in all experiments. 

    \item \textbf{Compare with cascade model serving system CascadeServe.} We compare \sys against an existing cascade model serving system CascadeServe. It chooses model cascade deployment plan based on system load (e.g., request arrival rate), enables model replication on hardware and adaptively dispatches incoming requests. We tune the parallelism and request routing strategies for CascadeServe based on the real-time system load and report the best values in all experiments.
\end{itemize}

\textbf{Traces.} We follow prior work to generate workload traces based on real-world data~\citep{jiang2024hexgen,zhong2024distserve}. Our testing traces are subsampled from MT-Bench~\citep{zheng2023judging}, a multi-turn conversation benchmark that contains multiple types of LLM workloads (e.g., coding, mathematics and reasoning). Each of our subsampled traces have different workload characteristics and different complexities as mentioned in~\S\ref{subsec:wt}.

\begin{figure}[t!]
    \centering
    \includegraphics[width=\linewidth]{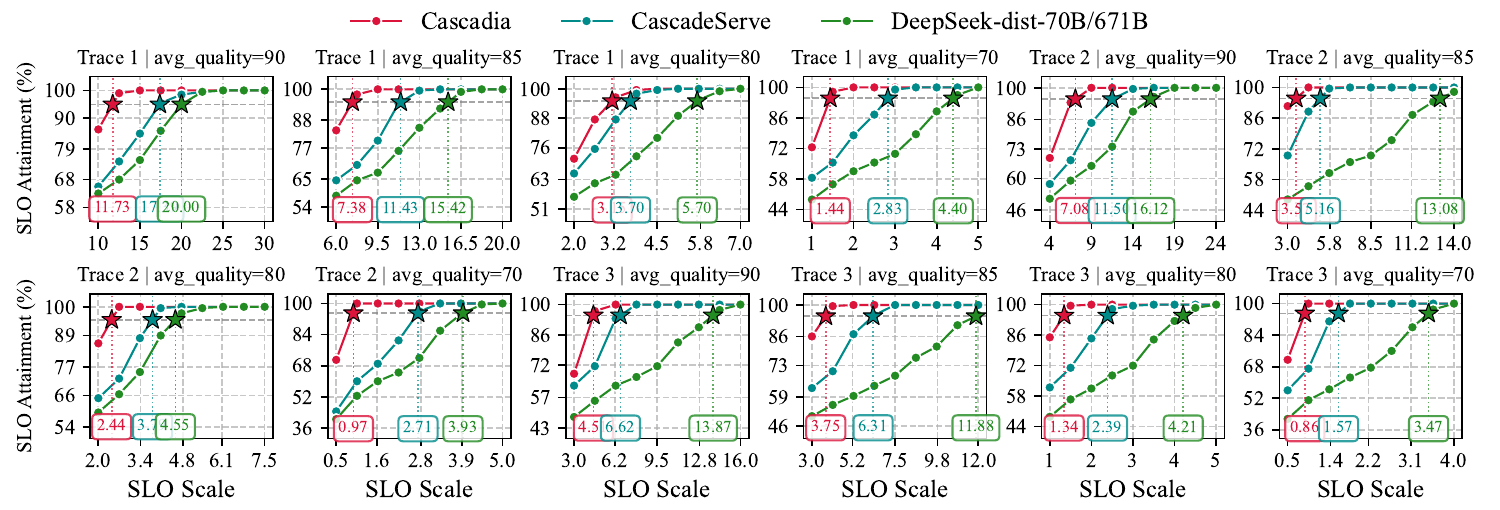}
    \caption{\small{End-to-end SLO attainment results evaluating \sys against two baseline systems. Each row corresponds to a particular LLM workload trace, and each column corresponds to a specific quality requirement. The stars indicate the 95\% SLO attainment for each system.}}
    \label{fig:slo}
\end{figure}

\textbf{Evaluation metrics.} Following previous evaluation setups~\citep{li2023alpaserve,duan2024muxserve,agrawal2024taming}, we evaluate system performance
based on SLO attainment and system throughput. The SLO is determined empirically based on the system's average single-request processing latency, and we scale it to various multiples (SLO Scale in~\autoref{fig:slo}) to assess
performance under different levels of operational stringency. We focus on identifying the minimum SLO Scale at which the system achieves 95\% SLO attainment.

\subsection{End-to-end Experimental Results}

\begin{figure}[t!]
    \centering
    \includegraphics[width=\linewidth]{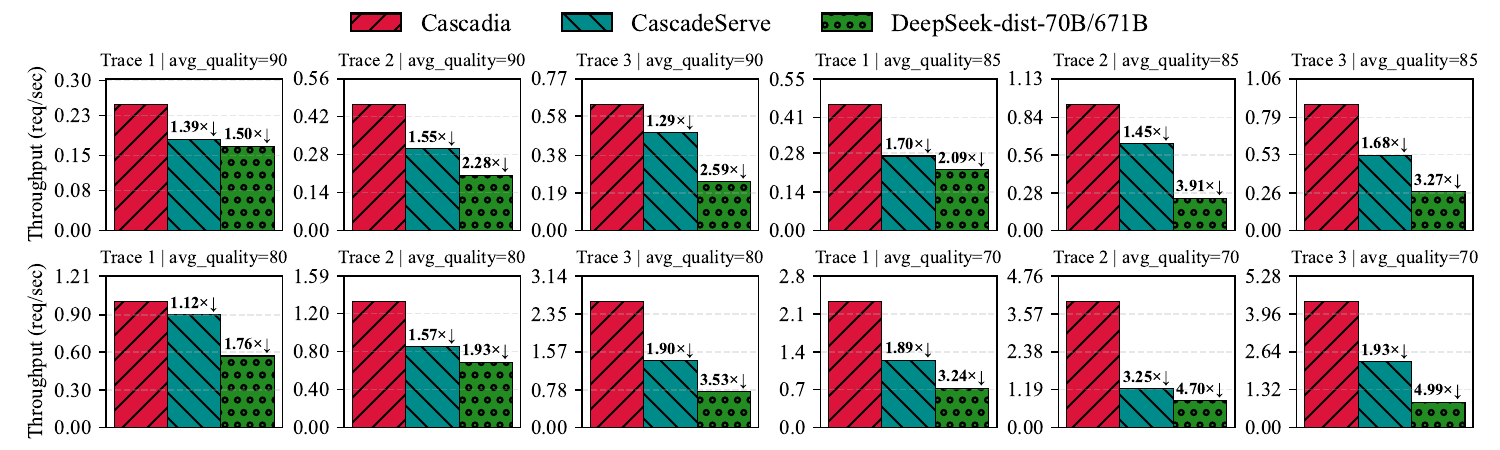}
    \caption{\small{End-to-end throughput results evaluating \sys against two baseline systems across different LLM workload traces and quality requirements.}}
    \label{fig:thpt}
\end{figure}

\textbf{End-to-end system performance.} We evaluate the SLO attainment and throughput of \sys across multiple traces and quality requirements, comparing it with two baselines. Results in~\autoref{fig:slo} and~\autoref{fig:thpt} show that \sys outperforms all baselines:
\begin{itemize}
    \item \sys achieves up to 4$\times$ and on average 2.8$\times$ lower latency deadlines, and up to 5$\times$ and on average 3$\times$ higher system throughput compared with stand-alone LLMs. For instance, when testing on trace 3 with an average quality requirement of 85, stand-alone DeepSeek-671B requires 11.88 SLO scales to achieve 95\% attainment, while \sys with different model types that uses smaller models to process simpler requests only requires 3.75 SLO scales.
\item \sys achieves up to 2.5$\times$ and on average 1.7$\times$ lower latency deadlines, and up to 3.3$\times$ and on average 1.7$\times$ higher throughput than CascadeServe. While CascadeServe optimizes model deployment and routing based on real-time load, it overlooks LLM-specific workload characteristics (e.g., input/output lengths) and request complexity, leading to sub-optimal parallelism and routing. For example, on trace 1 with an average quality requirement of 90, CascadeServe needs 17.3 SLO scales to reach 95\% SLO attainment, whereas \sys requires only 11.73.

\end{itemize}


\begin{figure}[t!]
    \centering
    \includegraphics[width=\linewidth]{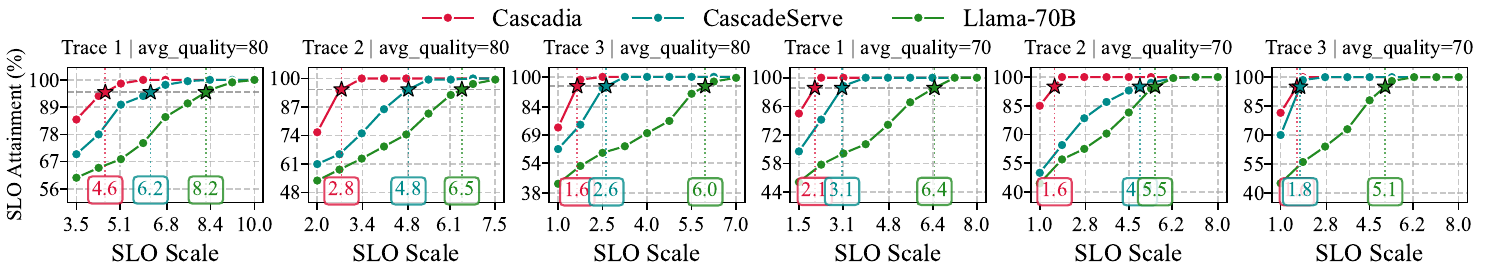}
    \caption{\small{End-to-end SLO attainment results evaluating \sys against two baselines using a Llama cascade (Llama3-8B; Llama3-70B) across LLM workload traces and quality requirements.}}
    \label{fig:others}
\end{figure}

\textbf{System performance with different model cascades \jyh{and serving optimizations.}} We further evaluate \sys using a different model cascade by replacing the DeepSeek series with the Llama series (Llama3-8B and Llama3-70B). As shown in~\autoref{fig:others}, \sys outperforms baselines by up to 3.8$\times$ and on average 2.6$\times$, demonstrating strong performance across LLM cascades. \jyh{We also compare \sys with Sarathi-Serve~\citep{agrawal2024taming}, a serving system with chunked prefill optimizations. \sys achieves 1.95$\times$ higher performance (1.64$\times$ average), validating our approach against advanced systems with scheduling optimizations. Detailed results are in~\autoref{appendix:sarathi-serve}.}

\subsection{Case studies on Model Deployment Plans and Routing Strategies}

\begin{wrapfigure}{r}{0.4\linewidth}
    \centering
    \includegraphics[width=\linewidth]{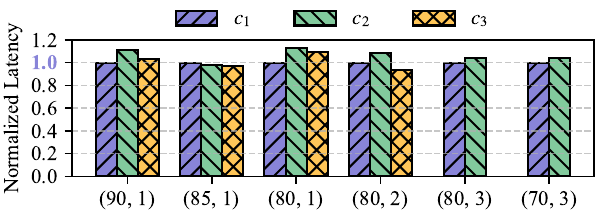}
    \caption{\small{Benchmarked p95 latency of each model type within the cascade across different testing cases.}}
    \label{fig:latency}
\end{wrapfigure}

\textbf{Case study on resource allocation and routing strategies.} We benchmarked the thresholds, processing ratios and allocated resources for different model types across different testing cases. For instance, when testing on trace 1 with an average quality requirement of 90, model types $c_1$ to $c_3$ process 100\%, 94\% and 50\% of the total requests, and the assigned GPU numbers are 4, 8 and 20. When the quality requirement changes to 85, less requests are required to be processed by the largest model $c_3$ (from 50\% to 21\%), and less resources are allocated to $c_3$ accordingly (from 20 to 16). This algorithm and system co-optimization enables \sys to adjust system resource allocation and request routing based on user requirements, ensuring balanced load across different model types to boost system performance. Additionally, when testing on trace 3 with an average quality requirement of 70, \sys deploys a subset of model types (DeepSeek-dist-7B and -70B) to minimize the latencies required for requests processing. As shown in~\autoref{fig:latency}, across different testing cases, \sys always balances the loads among different model types to ensure optimized system performance. \autoref{tab:case1} in \autoref{appendix:case study} demonstrates the thresholds, processing ratios and allocated resources for different model types across different testing cases.

\textbf{Case study on parallelism strategies.} We benchmarked the parallelism strategies for different model types across different testing cases. For example, when testing on trace 1 with an average quality requirement of 90, the optimal parallelism strategy $s_2$ for $c_2$ is (DP=2, TP=4). In this case, if we change the parallelism strategy to (DP=4, TP=2), the performance of this model type would drop by 33.7\%.
Additionally, when the quality requirement drops to 85, the optimal parallelism strategy $s_2$ for $c_2$ shifts to (DP=6, TP=2). This adjustment occurs because the change in quality requirements alters the LLM workloads, the request complexity routed and the resource allocated to $c_2$. Consequently, $s_2$ is updated to optimize the single model type’s performance while balancing loads across all model types within the cascade. \autoref{tab:case2} in \autoref{appendix:case study} presents the parallelism strategies for each model type within the cascade across different test cases.

\begin{figure}[t!]
    \centering
    \includegraphics[width=\linewidth]{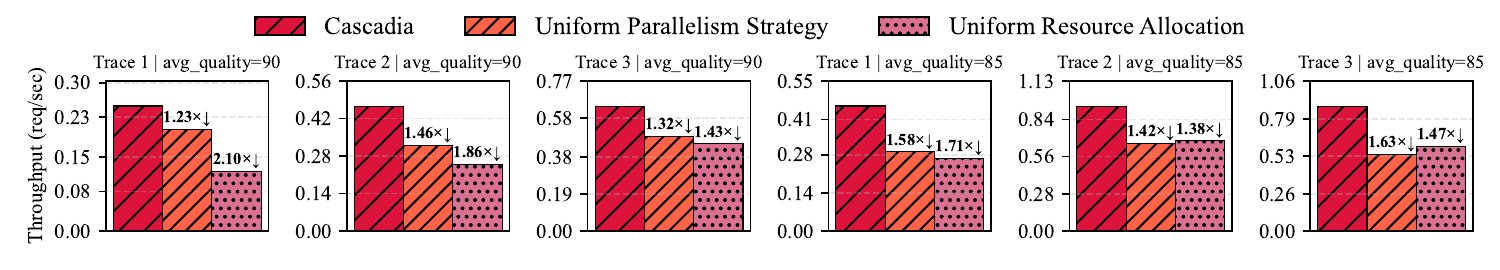}
    \caption{\small{Ablation study on resource allocation and parallelism strategy.}}
    \label{fig:abla}
\end{figure}

\textbf{Ablation study.} We disable individual optimizations in \sys to evaluate their impact, as shown in~\autoref{fig:abla}: (\textbf{\underline{i}}) Replacing our parallelism strategy optimization with a uniform parallelism strategy—tensor parallelism within each server and data parallelism across servers—reduces performance by up to 1.6$\times$ (1.4$\times$ on average). For example, DeepSeek‐7B and DeepSeek‐671B requires higher degrees of data and tensor parallelism to maximize throughput and parameter sharding; a uniform approach fails to accommodate these needs. (\textbf{\underline{ii}}) Replacing our resource allocation optimization with uniform resource allocation reduces performance by up to 2.1$\times$ (1.7$\times$ on average). For instance, in trace 1 with an average quality requirement of 90, DeepSeek‐671B was originally allocated 20 GPUs, but uniform allocation assigns only 12, causing load imbalance.

\subsection{Effectiveness of the Scheduling Algorithm}
\label{sec:re-scheduling}


\textbf{Overall scheduling process.} 
During scheduling, our Chebyshev-guided optimization (\S\ref{subsec:wt}) explores different routing strategies to reduce response latency given a required quality. Simultaneously, our MILP-based optimization (\S\ref{subsec:ilp}) searches for resource allocations and parallelism strategies to balance load across model types and minimize latency. \sys then selects the optimal plan—including thresholds, resource allocations, and parallelism strategies—based on quality requirements.



\begin{wrapfigure}{r}{0.4\linewidth}
    \centering
    \vspace{-1.25em}
    \includegraphics[width=\linewidth]{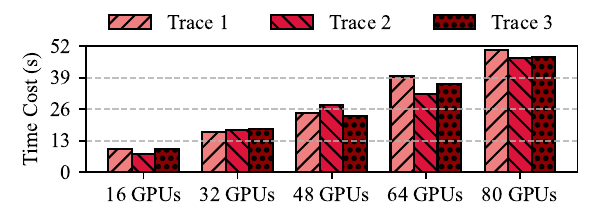}
    \vspace{-1.5em}
    \caption{\small{Algorithm running time when scaling from smaller clusters (e.g., 16 GPUs) to larger clusters (e.g., 80 GPUs).}}
    \vspace{-1em}
    \label{fig:time}
\end{wrapfigure}


\textbf{Scheduling algorithm runtime and scalability.} \autoref{fig:time} shows the runtime performance of \sys's scheduling algorithm, evaluated on a 12-core CPU instance. In our setup (32 GPUs), scheduling completes within 20s. For larger clusters (e.g., 80 GPUs), it finishes within one minute. These results demonstrate the algorithm’s efficiency and scalability across test cases and cluster sizes. Moreover, the algorithm is highly parallelizable, as resource allocations, parallelism, and routing strategies are independent—allowing execution time to scale down with more CPU cores.

\begin{wrapfigure}{r}{0.5\linewidth}
    \centering
    \includegraphics[width=\linewidth]{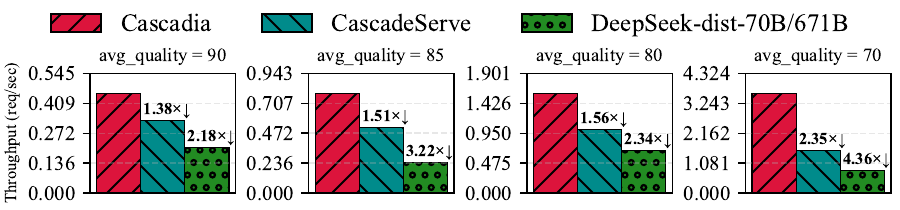}
    \caption{\small{Throughput evaluation under fluctuating workloads.}}
    \label{fig:rescheduling}
\end{wrapfigure}

\textbf{Re-scheduling to adapt to \jyh{online} workload changes.} 
As discussed in~\S\ref{subsec:wt}, LLM workload characteristics (e.g., \jyh{distributions of} input and output lengths, request rate and complexity) significantly affect the optimal model deployment plan and routing strategy. Thus, \jyh{analogous to DistServe~\citep{zhong2024distserve},} \sys implement a re-scheduling mechanism to accommodate dynamic LLM workloads. Concretely, the system (\textbf{\underline{i}}) subsample and record the real-time characteristics of the incoming LLM workloads (e.g., subsample 50 requests every 5 minutes and record the workload characteristics), (\textbf{\underline{ii}}) upon detecting a significant shift in workload characteristics (e.g., an increase in request arrival rate or request complexity), the scheduling algorithm is executed again, incorporating recent historical data to produce an updated deployment plan and routing strategy. 
\jyh{We evaluated our system against baselines under online fluctuating workloads, where the workload transitions trace 1 → trace 2 → trace 3 with segment lengths of 8, 16, and 10 minutes, evaluated at different quality constraints. As shown in~\autoref{fig:rescheduling}, \sys consistently outperforms baseline systems, achieving up to 4.4$\times$ improvement with an average of 2.2$\times$ better performance. Despite incurring additional scheduling overhead, \sys maintains superior throughput and end-to-end efficiency under fluctuating workloads by dynamically optimizing cascade plans based on real-time LLM workload characteristics.}

\section{Conclusion}
This paper proposes \sys, a cascade serving system tailored for LLMs. Its core component is a scheduling algorithm that jointly optimizes resource allocation, parallelism, and routing within the cascade system. Extensive experiments on diverse workload traces and multiple model cascades show that this co‑design substantially reduces request latency and boosts system throughput compared with both single‑model and existing cascade baselines, while maintaining the target answer quality.

\clearpage

\bibliographystyle{unsrt}
\bibliography{main}

\newpage
\appendix





\section{Extended Related Work}
\textbf{Parallelism strategies.} LLMs with huge memory and computational resource requirements typically rely on parallelization across multiple GPUs~\citep{li2023alpaserve,peng2025hexgen,jiang2022osdp,he2025efficient}. There are three prevalent forms of parallelism: data parallelism (DP, i.e., model replication), tensor parallelism (TP)~\citep{shoeybi2019megatron,yan2024hexiscale,miao2022galvatron}, and pipeline parallelism (PP)~\citep{huang2019gpipe,wang2024improving}.
DP replicates the model into multiple replicas, enabling parallel processing of requests. TP divides model weights and computationally intensive operations such as matrix multiplication across various GPUs, thereby splitting data scanning and computation to minimize LLM inference latency. PP divides the layers of a model into multiple stages. These stages are assigned to distinct GPUs for execution and they establish a pipeline. Only inter-layer activations are needed to be communicated between stages.

\textbf{Speculative decoding and early-exit in LLM inference.} Speculative decoding uses a lightweight draft model to generate token blocks, which a larger target model verifies—leveraging model heterogeneity to reduce computation and latency~\citep{leviathan2023fast,miao2024specinfer,liuonline}. Similarly, early-exit networks add decision branches at intermediate layers, enabling inference to stop early when confidence is high—cascading computation within a single model~\citep{teerapittayanon2016branchynet,rahmath2024early}. In contrast, we focus firmly on cascade model inference.

\section{Simulator Design and Validation}
\label{appendix:simu}

Our simulator employs a round-robin strategy for request dispatching among multiple parallel models, and a first-come first-served strategy for per-model request processing. The single-GPU processing time is based on profiled characteristics like compute TFLOPS and memory bandwidth. The simulator also considers the phase-specific characteristics of LLMs. The prefill phase is compute-bound, so its batched processing capacity is determined by the sum of the individual latencies. In contrast, the decoding phase is memory-bound, and its batched processing capability is defined by a single latency value. This distinction has been validated in several studies (e.g., DistServe~\citep{zhong2024distserve}, Splitwise~\citep{patel2024splitwise}).

\textbf{Inputs of the simulator.} The simulator requires three fundamental inputs: (i) the distributions of input and output sequence lengths for each model type within the cascade; (ii) the request arrival rate corresponding to each model type within the cascade; and (iii) the resource allocation designated for each model type within the cascade.

\textbf{Example.} Consider a workload distribution $\mathcal{W}$ that routes 100, 70, and 30 requests to model types 1, 2, and 3 respectively within the cascade, with corresponding GPU allocations of 2, 4, and 2 units. In this configuration, we record the distributions of input and output sequence lengths for each subset of requests (100, 70, and 30 respectively) as input files to the simulator, configure the request arrival rates and resource allocations according to the specified parameters, and execute the simulation. Subsequently, the simulator undergoes iterative execution to identify the optimal parallelism strategy based on the provided input files, request arrival rates, and resource allocation constraints.

\begin{table}[t!]
\centering
\caption{Simulator accuracy across parallelism configurations on Llama3-70B model under a workload with average input and output lengths of 1600 and 16. Errors are absolute percentage errors.}
\label{tab:sim-accuracy}
\begin{tabular}{lccc}
\hline
\textbf{Config (DP,TP,PP)} & \textbf{Real (req/s)} & \textbf{Estimated (req/s)} & \textbf{Abs.\ \% Error} \\
\hline
(1, 4, 1) & 0.21 & 0.219 & 4.29\% \\
\hline
(1, 2, 2) & 0.26 & 0.280 & 7.69\% \\
\hline
(1, 1, 4) & 0.27 & 0.287 & 6.30\% \\
\hline
(2, 1, 2) & 0.33 & 0.347 & 5.15\% \\
\hline
(2, 2, 1) & 0.40 & 0.408 & 2.00\% \\
\hline
(2, 4, 1) & 0.41 & 0.437 & 6.59\% \\
\hline
(2, 2, 2) & 0.55 & 0.559 & 1.64\% \\
\hline
\end{tabular}
\end{table}

\textbf{Batching strategy in our simulator.} The simulator's internal batching strategy is continuous batching, which iteratively batches request tokens to fully utilize the current resources. The GPU's memory limit constrains the maximum batch size for continuous batching.

\textbf{Queuing mechanism.} Our simulator maintains an individual queue for each model. Once there is free memory on the GPU (one request has finished), the model will fetch the next request in the queue for processing.

\textbf{Different parallelism.} Tensor and pipeline parallelism both split the computation workload of a single model across multiple devices. For pipeline parallelism, the simulator models communication overhead by profiling the relationship between estimated communication volume and observed latency. For tensor parallelism, the simulator assumes that each operator’s computation cost ideally scales down by a factor of $1/N$ when split across $N$ GPUs, and then adjusts this ideal cost using a speed-up coefficient $K(N)$ obtained from micro-benchmarks to account for communication and synchronization overhead. All profiling is performed offline before scheduling begins.

\textbf{Simulator evaluation.} We present the accuracy of our simulator with real-time experiments in~\autoref{tab:sim-accuracy}. The table presents examples of our throughput estimation for the Llama3-70B model under a workload with average input and output lengths of 1600 and 16, respectively. The notation (1,2,2) indicates a DP degree of 1, TP degree of 2, and PP degree of 2. Although the estimations are not perfectly accurate, they are sufficiently reliable (with estimation errors within 2\%–7\%) for selecting optimal configurations.

\section{Routing Solver in Latency-Constrained Case}
\label{appendix:routingsolver}

The routing solver can also optimize system response quality under a user-specified latency budget by solving
\[
\arg\min_{\theta}\left[ -\,Q(\theta)
\;+\; \nu\,\frac{\max\{0,\,L(\theta)-L_{\max}\}}{\,z^{\star}_{\mathrm{lat,\,max}}-z^{\star}_{\mathrm{lat,\,min}}\,} \right],
\]
where \(z^{\star}_{\mathrm{lat,\,min}}\) and \(z^{\star}_{\mathrm{lat,\,max}}\) are the best (minimum) and worst (maximum) achievable latencies, \(L_{\max}\) is the allowable latency budget, and \(\nu>0\) scales the penalty. The same routing–deployment alternation, deployment solver, and convergence procedure are reused unchanged.

\section{Complete Bi-level Optimization Formulation}
\label{appendix:bilevel}

\textbf{Problem setup and notation.}
We consider a cascade with $C$ model types/stages indexed by $\{1,\dots,C\}$ and labeled $\mathcal{C}=\{c_1,\dots,c_C\}$, where $c_i$ denotes the $i$-th model type.  
The routing strategy is denoted by $\theta$, parameterized by thresholds $\mathcal{H}=\{h_1,\dots,h_{C-1}\}$, with $\Theta$ the feasible set of routing strategies.  
The GPU resource allocation is $\mathcal{F}=\{f_1,\dots,f_C\}$, where $f_i\in\mathbb{Z}_+$ is the number of GPUs assigned to model type $i$, subject to a total budget $N\in\mathbb{Z}_+$.  
The parallelism plan is $\mathcal{S}=\{\mathrm{DP}_{i},\,\mathrm{TP}_{ij},\,\mathrm{PP}_{ij}\}_{i,j}$, where $\mathrm{DP}_i$ denotes the number of data-parallel replicas and, for each replica $j$, $\mathrm{TP}_{ij}$ and $\mathrm{PP}_{ij}$ denote its tensor- and pipeline-parallel degrees.  
Given routing $\theta$ and deployment $(\mathcal{F},\mathcal{S})$, the estimated p95 latency is $L(\theta,\mathcal{F},\mathcal{S})$, and the system quality is $Q(\theta;\tilde{\mathcal I})$ estimated by a judger using a subsampled workload $\tilde{\mathcal I}$.  
For Chebyshev-style normalization of quality, we use quality anchors $z_1^\star$ (utopia/best achievable quality, e.g., all requests at $c_C$) and $z_2^\star$ (nadir/worst credible quality, e.g., all requests at $c_1$).  
A user-specified quality requirement is $q_{\min}$, and $\mu>0$ is a penalty weight.

\textbf{Bi-level formulation.}
The routing is optimized by a single scalar objective that penalizes quality shortfall, normalized by the utopia–nadir range, while the deployment is optimized under the GPU budget and parallelism feasibility:
\[
\theta \in \operatorname*{arg\,min}_{\theta' \in \Theta}
\Bigg[
  L\!\big(\theta',\,\mathcal{F}^{\star},\,\mathcal{S}^{\star}\big)
  \;+\;
  \mu \,\max\!\left\{ 0,\; \frac{q_{\min}-Q(\theta';\,\tilde{\mathcal I})}{\,z_{1}^{\star}-z_{2}^{\star}\,} \right\}
\Bigg],
\]
\[
(\mathcal{F}^{\star},\mathcal{S}^{\star}) \in
\operatorname*{arg\,min}_{\mathcal{F},\,\mathcal{S}} L(\theta',\mathcal{F},\mathcal{S})
\quad \text{s.t.}\quad
\sum_{i=1}^{C} f_i \le N,\qquad
\sum_{j=1}^{\mathrm{DP}_i} \mathrm{TP}_{ij}\mathrm{PP}_{ij}=f_i\ \ (i{=}1,\dots,C),\qquad
\]
\[
f_i,\ \mathrm{DP}_i,\ \mathrm{TP}_{ij},\ \mathrm{PP}_{ij} \in \mathbb{Z}_+.
\]

\textbf{Tractability and solution strategy.}
Because the problem couples routing, resource allocation, parallelism, heterogeneous LLM workloads, and user-specific quality requirements, a monolithic solve is intractable.  
We therefore adopt a bi-level strategy: The deployment problem is solved as a MILP with latency values obtained from resource allocation and parallelism strategy optimization; the routing solver solves the Chebyshev-guided penalty problem.  
The two phases are executed iteratively, with the routing solver updating $\theta$ and the deployment solver resolving $(\mathcal{F}^{\star},\mathcal{S}^{\star})$ accordingly, and termination declared once the routing objective stabilizes under a prescribed horizon.

\textbf{Interpretation.}
The bi-level problem decomposes into \textbf{routing} and \textbf{deployment} subproblems that are solved iteratively.

\textbf{Deployment solver (deployment under resource/feasibility constraints).}
For a fixed routing $\theta'$, the deployment solver selects the latency-optimal deployment by choosing GPU allocations and parallelism plans subject to the budget and structural constraints:
\[
(\mathcal{F}^{\star},\mathcal{S}^{\star}) \in
\operatorname*{arg\,min}_{\mathcal{F},\,\mathcal{S}} L(\theta',\mathcal{F},\mathcal{S})
\quad \text{s.t.}\quad
\sum_{i=1}^{C} f_i \le N,\qquad
\sum_{j=1}^{\mathrm{DP}_i} \mathrm{TP}_{ij}\mathrm{PP}_{ij}=f_i\ \ (i{=}1,\dots,C),\qquad
\]
\[
f_i,\ \mathrm{DP}_i,\ \mathrm{TP}_{ij},\ \mathrm{PP}_{ij} \in \mathbb{Z}_+.
\]
This solver captures both hardware limits (GPU budget $N$) and parallelism feasibility.

\textbf{Routing solver (routing, Chebyshev-guided optimization).}
Given the current deployment $(\mathcal{F}^{\star},\mathcal{S}^{\star})$, the routing solver updates the routing strategy (i.e., $\theta$) by minimizing a single scalar objective that balances latency and a normalized quality shortfall:
\[
\theta \in \operatorname*{arg\,min}_{\theta' \in \Theta}
\Bigg[
  L\!\big(\theta',\,\mathcal{F}^{\star},\,\mathcal{S}^{\star}\big)
  \;+\;
  \mu \,\max\!\left\{ 0,\; \frac{q_{\min}-Q(\theta';\,\tilde{\mathcal I})}{\,z_{1}^{\star}-z_{2}^{\star}\,} \right\}
\Bigg].
\]
Here, $(z_{1}^{\star}\!-\!z_{2}^{\star})^{-1}$ provides Chebyshev (utopia–nadir) normalization for scale stability, and $\mu>0$ sets the severity of penalizing $Q(\theta')<q_{\min}$. For sufficiently large $\mu$ (when the target is feasible), any minimizer is quality-compliant and the routing objective effectively reduces to minimizing latency among feasible routings.

\textbf{Coupling and procedure.}
The routing solver’s $\theta$ determines the workload distribution seen by each model type within the cascade (and hence the optimal deployment plan for the deployment solver), while the deployment solver’s $(\mathcal{F}^{\star},\mathcal{S}^{\star})$ determines the latency used by the routing objective (and hence the optimal routing strategy for the routing solver). Alternating updates continue until the routing objective stabilizes under a prescribed termination horizon (e.g., best-so-far objective unchanged for $K$ consecutive iterations).

\section{Case studies on Model Deployment Plans and Routing Strategies}
\label{appendix:case study}

\textbf{Case study on resource allocation and routing strategies.}
\autoref{tab:case1} demonstrates the case study of thresholds, processing ratios and allocated resources for different model types across different testing cases.

\begin{table}[htbp]
\centering
\caption{\small{Case study of the thresholds ($h_1$, $h_2$), processing ratios ($p_1$, $p_2$, $p_3$), and allocated resources ($f_1$, $f_2$, $f_3$) for each model type within the cascade across different testing cases. (90, 1) denotes testing on Trace 1 with an average quality requirement of 90.}}
\resizebox{0.7\textwidth}{!}{
\begin{tabular}{c >{\centering\arraybackslash}p{1cm} >{\centering\arraybackslash}p{1cm} >{\centering\arraybackslash}p{1cm} >{\centering\arraybackslash}p{1cm} >{\centering\arraybackslash}p{1cm} >{\centering\arraybackslash}p{1cm} >{\centering\arraybackslash}p{1cm} >{\centering\arraybackslash}p{1cm}}
\hline
& \textbf{$h_1$} & \textbf{$h_2$} & \textbf{$p_1$} & \textbf{$p_2$} & \textbf{$p_3$} & \textbf{$f_1$} & \textbf{$f_2$} & \textbf{$f_3$} \\
\hline
(90, 1) & 99 & 91 & 100\% & 94\% & 50\% & 4 & 8 & 20 \\
\hline
(85, 1) & 74 & 64 & 100\% & 62\% & 21\% & 4 & 12 & 16 \\
\hline
(80, 1) & 69 & 25 & 100\% & 54\% & 11\% & 6 & 14 & 12 \\
\hline
(80, 2) & 61 & 18 & 100\% & 31\% & 3\% & 8 & 16 & 8 \\
\hline
(80, 3) & 32 & 0 & 100\% & 23\% & 0\% & 18 & 14 & 0 \\
\hline
(70, 3) & 10 & 0 & 100\% & 5\% & 0\% & 24 & 8 & 0 \\
\hline
\end{tabular}
}
\label{tab:case1}
\end{table}

\textbf{Case study on parallelism strategies.} \autoref{tab:case2} presents a case study on parallelism strategies for each model type within the cascade across different test cases.

\begin{table}[htbp]
\centering
\caption{\small{Case study of the parallelism strategies for each model type within the cascade ($s_1$, $s_2$, $s_3$) across different testing cases.}}
\resizebox{0.7\textwidth}{!}{
\begin{tabular}{c c}
\hline & \textbf{Parallelism Strategies} \\
\hline
(90, 1) & $s_1$: (DP=4), $s_2$: (DP=2, TP=4), $s_3$: (TP=4, PP=3), (TP=8) \\
\hline
(85, 1) & $s_1$: (DP=2, TP=2), $s_2$: (DP=6, TP=2), $s_3$: (DP=2, TP=8) \\
\hline
(80, 1) & $s_1$: (DP=6), $s_2$: (DP=5, TP=2), (TP=4), $s_3$: (TP=4, PP=3) \\
\hline
(80, 2) & $s_1$: (DP=6), (TP=2), $s_2$: (DP=8, TP=2), $s_3$: (TP=8) \\
\hline
(80, 3) & $s_1$: (DP=10), (DP=4, TP=2), $s_2$: (DP=2, TP=4), (DP=3, TP=2), $s_3$: - \\
\hline
(70, 3) & $s_1$: (DP=16), (DP=4, TP=2), $s_2$: (DP=4, TP=2), $s_3$: - \\
\hline
\end{tabular}
}
\label{tab:case2}
\end{table}

\begin{table}[htbp]
\centering
\caption{End-to-end throughput results evaluating \sys against Sarathi-Serve.}
\label{tab:throughput}
\begin{tabular}{lcccc}
\hline
\textbf{Trace} & \textbf{Ours} & \textbf{Sarathi-Serve} & \textbf{Speedup} & \textbf{\%Improvement} \\
\hline
Trace 1 & 0.2529 req/s & 0.1913 req/s & 1.322 & +32.20\% \\
Trace 2 & 0.4659 req/s & 0.2385 req/s & 1.953 & +95.35\% \\
Trace 3 & 0.6406 req/s & 0.3977 req/s & 1.611 & +61.08\% \\
\hline
\end{tabular}
\end{table}

\section{Comparison with Sarathi-Serve}
\label{appendix:sarathi-serve}

We evaluated Sarathi-Serve under the same experimental setup as SGLang, as described in~\S\ref{sec:experimental_setup}, using traces 1–3 with an average quality requirement of 90. We used Sarathi-Serve's vLLM implementation (its most efficient variant) and tuned the chunk size to be optimal for each case. As shown in~\autoref{tab:throughput}, our system achieves up to 1.95$\times$ higher throughput and averages a 1.64$\times$ speedup across traces.

\end{document}